\documentclass{appolb}

\usepackage{epsfig}

\begin{document}

\title{
The Shears Mechanism in $^{\rm 142}$Gd \\
in the Skyrme-Hartree-Fock Method \\
with the Tilted-Axis Cranking%
\thanks{
Presented at the XXVII Mazurian Lakes School of Physics, September 2-9 2001, Krzy\.ze, Poland}%
}

\author{
P. Olbratowski,$^{\rm a,b}$ J. Dobaczewski,$^{\rm a}$ J. Dudek,$^{\rm b}$ \\
T. Rz{\c a}ca-Urban,$^{\rm c}$ Z. Marcinkowska,$^{\rm c}$ and R.M. Lieder$^{\rm d}$
\address{
$^{\rm a}$Institute of Theoretical Physics, Ho\.za 69, 00-681 Warsaw, Poland\\
$^{\rm b}$Institut de Recherches Subatomiques, 23 rue du Loess, 67037 Strasbourg, France\\
$^{\rm c}$Institute of Experimental Physics, Ho\.za 69, 00-681 Warsaw, Poland\\
$^{\rm d}$Institut f\"ur Kernphysik, Foschungszentrum J\"ulich, 52425 J\"ulich, Germany}}

\maketitle

\begin{abstract}
We report on the first Skyrme-Hartree-Fock calculations with the tilted-axis cranking in the context of magnetic rotation. The mean field symmetries, differences between phenomenological and self-consistent methods and the generation of shears-like structures in the mean field are discussed. Significant role of the time-odd spin-spin effective interaction is pointed out. We reproduce the shears mechanism, but quantitative agreement with experiment is rather poor. It may have to do with too large core polarization, lack of pairing correlations or properties of the Skyrme force.
\end{abstract}


\section{Introduction}

It is known, that if a quantum system exhibits rotational excitations
there must be a factor breaking its spherical symmetry. In the
nuclear case, deformation of the charge distribution is the most
familiar one, and was the only known till the early 1990's. In that
period a new situation was found in some rotational bands of light
lead isotopes. These bands are characterized by very weak E2
intraband transitions, implying almost spherical shape, and strong M1
transitions, suggesting large magnetic dipole moment as the main
factor breaking sphericity. This is why the new phenomenon was given
the name of {\it magnetic rotation}, see Ref.~\cite{[Fra00]} for a
review.

The following mechanism is believed to underly the so-called {\it
magnetic dipole bands}. At the bandhead, the spin of the valence
protons and that of the valence neutrons form the angle of
90$^\circ$. When the system rotates, both spins align towards the axis of
rotation due to the gyroscopic effect, resembling the closing of
shears blades. Hence the name of the {\it shears mechanism}.

The cranked mean-field is a tool suitable for describing rotational
excitations. There is a well established correspondence between the
spin-parity sequences in rotational bands, and the discrete
symmetries of the mean field \cite{[Boh75]}. Properties of the
standard E2 bands imply conservation of parity and signature. With
these symmetries imposed, the cranking frequency can be applied only
along the axis to which the signature refers, and all average-spin
vectors obtained from the solution point along this unique direction.
Such a case is called the {\it one-dimensional} or {\it
principal-axis cranking}. For the magnetic dipole bands, the only
conserved spatial symmetry is the parity. In the context of the mean
field, this allows for any direction of the cranking-frequency and
average-spin vectors which complies with the perpendicular
orientation in the shears picture. Such an approach is referred to as
the {\it tilted-axis cranking}.

There are two basic versions of the mean field approach -
phenomenological, employing a pre-defined potential, and
self-consistent, in which the potential is obtained from averaging a
two-body effective interaction. Up to now, only the former one has
been used for tilted-cranking calculations, with the notable
exception of a self-consistent calculation performed within the
relativistic mean field applied to $^{84}$Rb \cite{[Mad00]}. In the
phenomenological method, one can either fix the orientation of the
potential and vary the direction of the rotational frequency vector
when minimizing energy or vice-versa. The first possibility is used
in existing codes. Self-consistently, the orientation of the
potential is not available as a parameter, and the only feasible way
is to fix the cranking vector and let the self-consistent potential
reorient and conform to it in course of the Hartree-Fock iterations.

Although the phenomenological description of the shears bands has led
to a considerable success, the self-consistent methods have to be
applied in order to study such structures in more detail. This
includes the stability of the proposed configurations with respect to
the core excitation, the full minimization of the underlying energies
with respect to all deformation variables, and the inclusion of the
spin-current interactions.

\section{The code HFODD}

In the present study, the self-consistent solutions for rotational
bands in $^{142}$Gd were obtained with the new version (v1.91) of the
code HFODD \cite{[Olb02]}. Previous versions are described in
Ref.~\cite{[Dob00d]}. The code solves the nuclear Hartree-Fock
equations for the Skyrme effective interaction. All time-even and
time-odd terms of the mean field can be included. The wavefunctions
are expanded onto the deformed Cartesian harmonic-oscillator basis.
Four symmetry modes are available, conserving either both parity and
signature, only simplex, only parity, or no point symmetries. The
cranking-frequency vector can take any direction. If the solution
becomes tilted with respect to the laboratory frame, the program
finds the principal axes of the mass quadrupole moment and
recalculates all quantities in the corresponding intrinsic frame.
Pairing correlations are not taken into account.

\section{Results obtained for $^{\rm 142}$Gd}

It turns out, that the Skyrme-Hartree-Fock calculations with broken
signature are very vulnerable to divergences and often parallel
coupling of angular momenta is obtained where a perpendicular one is
expected. We attribute these effects to overestimated strengths of
the time-odd components of the Skyrme mean field. Terms originating
from the interaction between intrinsic spins have a particularly
strong influence here. This observation complies with what is known
about the nuclear spin-spin interaction, that it preferrers
parallel coupling. The terms in question are $\vec{s}^{~2}$ and
$\vec{s}\cdot\Delta\vec{s}$ - see Refs.~\cite{[Dob95e],[Dob00d]} for
details. We tested the SKM* \cite{BQB82} and SLy4 \cite{[Cha97]}
forces and observed a similar behavior in both cases. In the present
calculations the SLy4 parametrization was used and the time-odd
coupling constants corresponding to $\vec{s}^{~2}$ (both density
dependent and independent), $\vec{s}\cdot\Delta\vec{s}$ and
$\vec{s}\cdot\vec{T}$ were set to zero. Only the ones multiplying the
terms $\vec{j}^{~2}$ and $\vec{s}\cdot(\nabla\times\vec{j})$ were kept at
the original values of the SLy4 parametrization, in order to
conserve the local gauge invariance of the force \cite{[Dob95e]}.

The nucleus $^{\rm 142}$Gd ($Z$=64 and $N$=78) belongs to the
so-called transitional region of nuclei, i.e., it has a small but
non-negligible deformation $\beta$ and a substantial triaxiality
angle $\gamma$. It may, therefore, exhibit an interplay between the
shears mechanism and the standard collective rotation. Fig.~1 shows a
partial level scheme of $^{\rm 142}$Gd taken from
Ref.~\cite{[Lie01]}. Bands denoted by $\pi h_{11/2}^{2}$ and $\nu
h_{11/2}^{-2}$, as well as the ground-state band, have the electric
quadrupole character. Band denoted by $\pi h_{11/2}^{2}~\nu h_{11/2}^{-2}$ is a
magnetic dipole one.

\begin{figure}
\begin{center}
\leavevmode
\epsfxsize=\textwidth \epsfbox{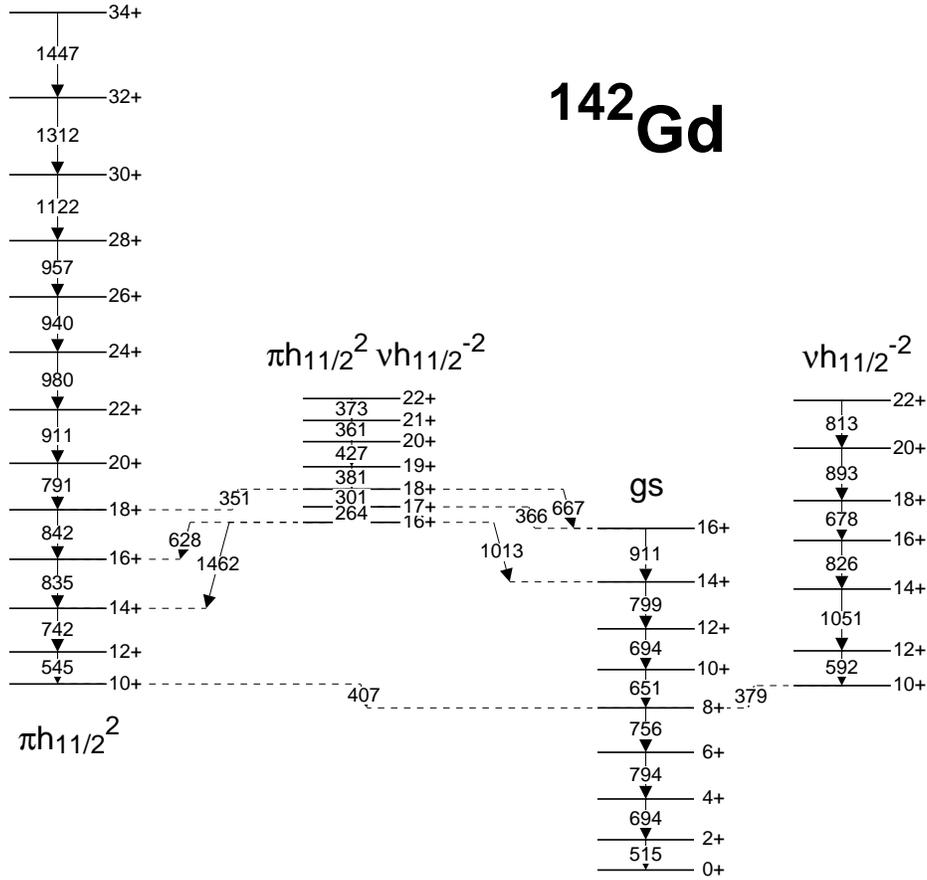}
\end{center}
\caption{Part of the $^{\rm 142}$Gd level scheme from Ref.~\protect\cite{[Lie01]}}
\end{figure}

In our calculations all the four bands correspond to the same
slightly triaxial deformation of about $\beta=0.19$ and
$\gamma=40^\circ$, that does not significantly change with rotational
frequency. It is known, that spherical mean-field orbitals split with
deformation. For oblate shapes the suborbitals with high Nilsson
number $\Omega$ (angular momentum projection onto the symmetry axis)
become lower in energy than those with low $\Omega$. Our solution is
close to the oblate shape and we observe a similar behavior. In
particular, the lowest suborbitals of $h_{11/2}$ have large
alignments along the shortest axis of the nucleus and small
alignments along the longest one. The opposite holds for the highest
suborbitals.

\begin{figure}
\begin{center}
\leavevmode
\epsfxsize=\textwidth \epsfbox{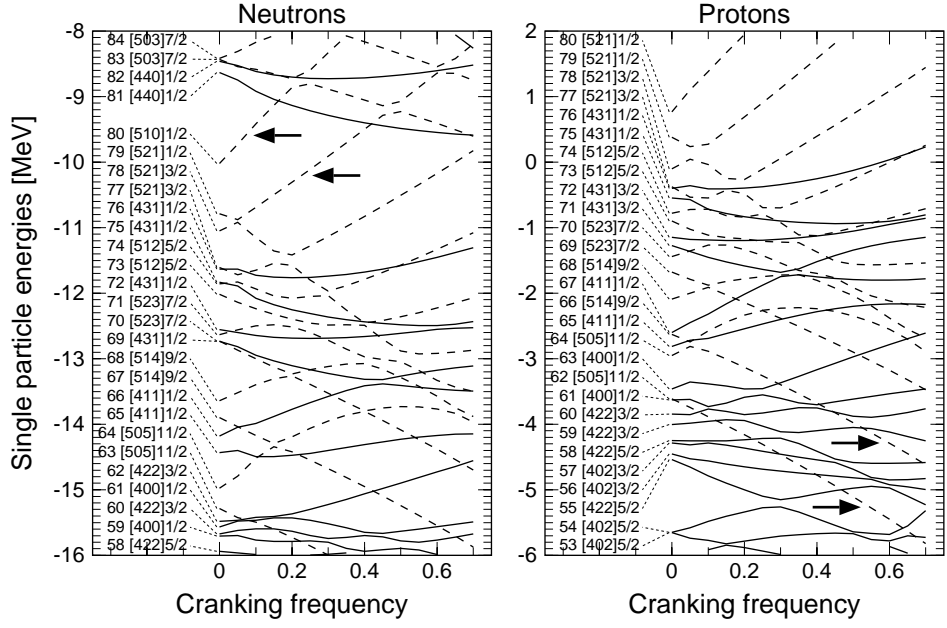}
\end{center}
\caption{Single-particle routhians in  $^{\rm 142}$Gd calculated for
the  $\pi h_{11/2}^{2}~\nu h_{11/2}^{-2}$ configuration. The two
proton and two neutron routhians responsible for the shears blades
are marked by arrows.}
\end{figure}

In the ground state obtained from our calculations there are two
protons in $h_{11/2}$ occupying its two lowest suborbitals. They
carry alignments of about $\pm$11/2 along the shortest axis, and the
total equals zero. When either of them is excited to the state of
alignment $\mp$9/2, spin of 10 units along the shortest axis is
obtained. This corresponds to the bandhead of the band denoted as
$\pi h_{11/2}^{2}$ in Fig.~1. By applying the cranking field along
this direction one obtains the subsequent states of the band.
Analogous excitation of neutron holes yields spin 10, but along the
longest axis, because the holes are located in the highest
suborbitals of $h_{11/2}$. This leads to the band marked as $\nu
h_{11/2}^{-2}$.

\begin{figure}
\begin{center}
\leavevmode
\epsfxsize=10cm \epsfbox{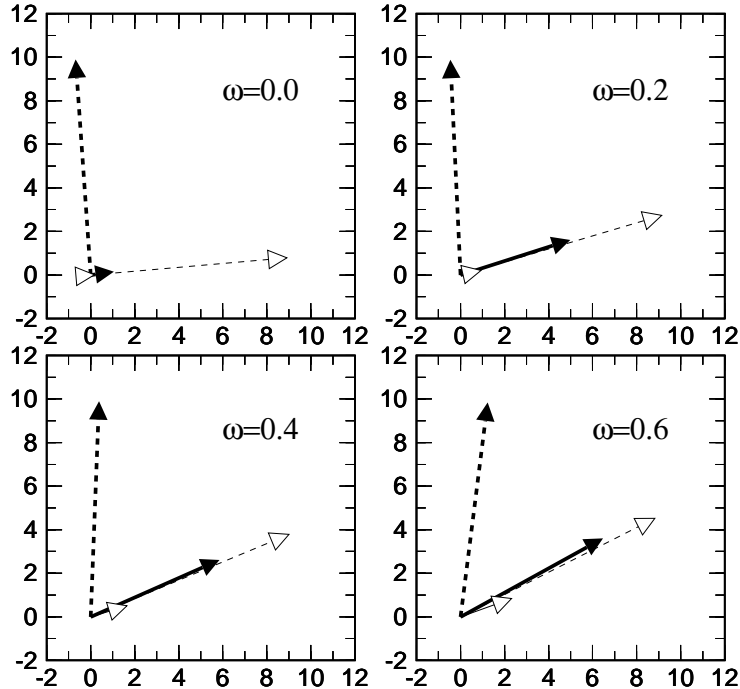}
\end{center}
\caption{Angular momentum vectors in the plane spanned by the
shortest and longest axes of the nucleus. Proton and neutron
contributions are marked by the thick lines with full arrowheads and
thin lines with open arrowheads, respectively. Dashed lines denote
the contributions from the valence particles in $h_{11/2}$, that is,
the shears blades. Solid lines refer to the core subsystems.
Rotational frequency $\omega$ is given in units of MeV/$\hbar$.}
\end{figure}

When the neutron and proton excitations are combined together,
perpendicular coupling is obtained, corresponding to the magnetic
dipole band $\pi h_{11/2}^{2}~\nu h_{11/2}^{-2}$. The single-particle
routhians calculated for this band are shown in Fig.~2. Note, that
they behave quite differently than in the one-dimensional cranking,
namely, all components of $h_{11/2}$ split strongly with rotational
frequency. The proton particles and neutron holes responsible for the
two shears blades are marked by arrows.

Fig.~3 shows the angular momentum vectors for several cranking
frequencies. Spins of the $h_{11/2}$ protons and neutrons, that is,
the shears blades (dashed lines), and contributions from the
remaining proton and neutron cores (solid lines) are shown
separately. Closing of the shears with rotational frequency is
clearly visible. At the same time, polarization of the core increases
rapidly. As expected, both cores polarize in the direction of the
longest axis, which plays the role of the collective one, because the
nucleus is approximately oblate. The angular momentum of the proton
core is much larger than that of the neutron one. The two core
subsystems differ by the six $d_{5/2}$ neutrons, so the properties of
this orbital may be responsible for the obtained asymmetry.

Experimentally, core polarization in $^{\rm 142}$Gd certainly affects
the properties of the $\pi h_{11/2}^{2}~\nu h_{11/2}^{-2}$ band - it
is observed up to spin 22, which is by 2 units more than the two
blades of spin 10 can account for. Similar core-rotation effects have
already been considered in the phenomenological analysis of the
$^{197}$Pb shears band \cite{[Coo01]}. It seems, however, that the
core polarization is too strong in our calculations, because
we can build the band far beyond the spin 22 without even achieving
zero angle between the shears. For example, solution for
$\omega$=0.6\,MeV/$\hbar$ illustrated in Fig.~3, corresponds to
$I$$\simeq$26. Another point is that the calculated moment of inertia
is too large, see Fig.~4.

\begin{figure}
\begin{center}
\leavevmode
\epsfxsize=9cm \epsfbox{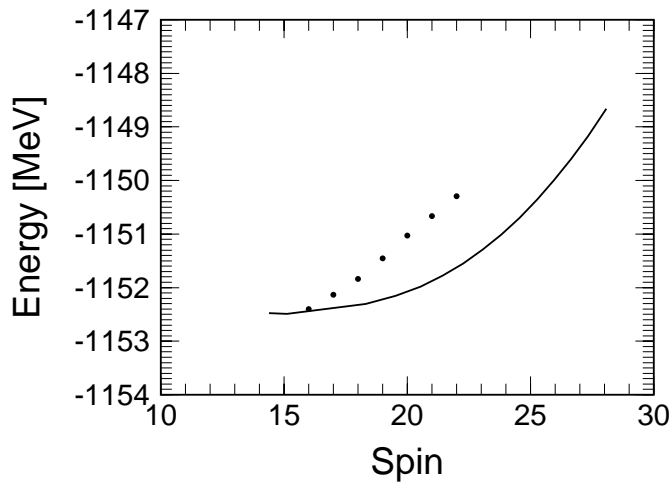}
\end{center}
\caption{Energy as a function of spin for the magnetic dipole band
$\pi h_{11/2}^{2}~\nu h_{11/2}^{-2}$ in $^{\rm 142}$Gd. Vertical
scale shows the total energy obtained from the present calculations
(line). Experimental data (dots) are arbitrarily shifted in energy in
order to make the point at spin 16 coincide with the line. }
\end{figure}

\section{Conclusions}

In the present paper we report on the first Skyrme-Hartree-Fock
calculations performed within the tilted-axis cranking method. It
seems, that the strength of the time-odd components of the Skyrme
mean field is crucial for the existence of the shears-like solutions,
and that these strengths are incorrect when taken directly from
typical Skyrme-force parametrizations. Our results for the $\pi
h_{11/2}^{2}~\nu h_{11/2}^{-2}$ band in $^{\rm 142}$Gd indicate, that
both the shears mechanism and collective rotation of the core are
responsible for the generation of angular momentum, although the
latter effect is probably too strong in the calculations. Whether the
reason for the obtained discrepancies is the lack of pairing
correlations or inadequate time-odd terms is not clear at the moment,
and will be the subject of further investigations.

\section{Acknowledgments}
This work was supported in part by the Polish Committee for
Scientific Research (KBN) under Contract No.~5~P03B~014~21 and by the
French-Polish integrated actions program POLONIUM.


\end{document}